\documentclass[12pt]{article}
\usepackage[dvips]{graphicx}
\def\be{\begin{equation}}					 
\def\ee{\end{equation}}
\def\ber{\begin{eqnarray}}
\def\eer{\end{eqnarray}}	

\begin{document}
\vspace*{1cm}
\begin{center}
{\Large \bf Kaluza-Klein Picture and Mass Spectrum \\[1ex] of
Two-Pion System}\\

\vspace{4mm}

{\large A.A. Arkhipov\\
{\it State Research Center ``Institute for High Energy Physics" \\
 142280 Protvino, Moscow Region, Russia}}\\
\end{center}

\vspace{4mm}
\begin{abstract}
{In this note we present additional arguments in favour of
Kaluza and Klein picture of the world. In fact, we have shown that
formula (\ref{KKpi}) provided by Kaluza-Klein approach with the
fundamental scale early calculated (Eq.~\ref{scale} \cite{1})  gives
an excellent description for the mass spectrum of two-pion system.} 
\end{abstract}

\section{Introduction}

In our previous papers \cite{1,2} we have presented the arguments in
favour of that the Kaluza-Klein picture of the world has been been
observed in the experiments at very low energies where the
nucleon-nucleon dynamics has been studied. In particular we have
found that geniusly simple formula for KK excitations provided by
Kaluza-Klein approach gives an excellent description for the mass
spectrum of two-nucleon system. Surely, this was quite an event and,
certainly, this very nice fact inspired us to study the other
two-particle hadronic systems in this respect.

Recently there is a large amount of data concerning resonance states
of two-pion system \cite{3,4}. The modern strong interaction theory
formulated in terms of known QCD Lagrangian do not allow us at
present time to make an appreciable breakthrough in the problem of
calculating the masses of compound systems mainly because this
problem is a significantly non-perturbative one. Therefore, great
efforts have been undertaken to develop various kinds
(semi)phenomenological models for systematics of experimentally
observed resonance states.

We have performed an analysis of experimental data on mass spectrum
of the resonance states of two-pion system and compared them with
Kaluza-Klein picture. The results of this analysis have
been presented here.  

\section{Kaluza-Klein picture and KK excitations in two-pion
system}

It is well known that the basic idea of the Kaluza-Klein scenario may
be applied to any model in Quantum Field Theory (see for the details
e.g. the excellent review articles \cite{5,6} and many references
therein). As example, let us consider the simplest case of
(4+d)-dimensional model of scalar field with the action  
\be
S = \int d^{4+d}z \sqrt{-{\cal G}} \left[
\frac{1}{2} \left( \partial_{M} \Phi \right)^2 - 
\frac{m^{2}}{2} \Phi^2 + \frac{G_{(4+d)}}{4!} \Phi^4
\right], 
\label{S}
\ee
where ${\cal G}=\det|{\cal G}_{MN}|$, ${\cal G}_{MN}$ is the metric
on ${\cal M}_{(4+d)} = M_4 \times K_d$, $M_4$ is pseudo-Euclidean
Minkowski space-time, $K_d$ is a compact internal $d$-dimensional
space with the characteristic size $R$. Let $\Delta_{K_{d}}$ be the
Laplace operator on the internal space $K_{d}$, and $Y_{n}(y)$ are
ortho-normalized eigenfunctions of the Laplace operator 
\be
\Delta_{K_{d}} Y_{n}(y) = -\frac{\lambda_{n}}{R^{2}} Y_{n}(y),  
\label{Yn}
\ee
and $n$ is a (multi)index labeling the eigenvalue
$\lambda_{n}$ of the eigenfunction $Y_{n}(y)$. $d$-dimensional torus
$T^{d}$ with equal radii $R$ is an especially simple example of the
compact internal space of extra dimensions $K_d$. The eigenfunctions
and eigenvalues in this special case look like 
\be
Y_n(y) = \frac{1}{\sqrt{V_d}} \exp \left(i \sum_{m=1}^{d}
n_{m}y^{m}/R
\right), \label{T}
\ee
\[
\lambda_n = |n|^2,\quad |n|^2= n_1^2 + n_2^2 + \ldots n_d^2, \quad
n=(n_1,n_2, \ldots, n_d),\quad -\infty \leq n_m \leq \infty,
\]
where $n_m$ are integer numbers, $V_d = (2\pi R)^d$ is the
volume of the torus.

To reduce the multidimensional theory to the effective
four-dimensional one we wright a harmonic expansion for
the multidimensional field $\Phi(z)$ 
\be
\Phi(z) = \Phi(x,y) = \sum_{n} \phi^{(n)}(x) Y_{n}(y). 
\label{H}
\ee
The coefficients $\phi^{(n)}(x)$ of the harmonic expansion
(\ref{H}) are called Kaluza-Klein (KK) excitations or KK modes, and
they usually include the zero-mode $\phi^{(0)}(x)$, corresponding to
$n=0$ and the eigenvalue $\lambda_{0} = 0$. Substitution of the KK
mode expansion into action (\ref{S}) and integration over the
internal space $K_{d}$ gives
\be
S = \int d^{4}x \sqrt{-g} \left\{  
\frac{1}{2} \left( \partial_{\mu} \phi^{(0)} \right)^{2} -
\frac{m^{2}}{2}
(\phi^{(0)})^{2} \right. + \frac{g}{4!} (\phi^{(0)})^{4} +
\ee
\[
+\left. \sum_{n \neq 0} \left[\frac{1}{2}
\left(\partial_{\mu} \phi^{(n)}
\right) 
\left(\partial^{\mu} \phi^{(n)} \right)^{*} -\frac 
{m_n^2}{2} \phi^{(n)}\phi^{(n)*} \right] 
+ \frac{g}{4!} (\phi^{(0)})^{2} \sum_{n\neq 0} \phi^{(n)}
\phi^{(n)*}\right\} + \ldots.  
\]
For the masses of the KK modes one obtains 
\be
m_{n}^{2} = m^{2} + \frac{\lambda_{n}}{R^{2}}, \label{m}
\ee
and the coupling constant $g$ of the four-dimensional theory is
related  to the coupling constant $G_{(4+d)}$ of the initial
multidimensional theory by the equation 
\be
  g = \frac{G_{(4+d)}}{V_d},  \label{g}
\ee
where $V_d$ is the volume of the compact internal space of extra
dimensions $K_d$. The fundamental coupling constant $G_{(4+d)}$ has
dimension $[\mbox{mass}]^{-d}$. So, the four-dimensional coupling
constant $g$ is dimensionless one as it should be.
Eqs.~(\ref{m},\ref{g}) represent the basic
relations of Kaluza-Klein scenario.  Similar relations take place for
other types of multidimensional quantum field theoretical models.
From four-dimensional point of view we can interpret each KK mode as
a particle with the mass $m_n$ given by Eq.~(\ref{m}). We see that in
according with Kaluza-Klein scenario any multidimensional
field contains an infinite set of KK modes, i.e. an infinite set of
four-dimensional particles with increasing masses, which is called
the Kaluza-Klein tower. Therefore, an experimental observation of
series KK excitations with a characteristic spectrum of the form
(\ref{m}) would be an evidence of the existence of extra dimensions.

We have applied the main issues of Kaluza-Klein approach to our
analysis of the structure of proton-proton total cross section at
very low energies and calculated the fundamental scale (size) $R$ of
the compact internal extra space. One obtained by this way
\be
\frac{1}{R} = 41.481\,\mbox{MeV}\quad \mbox{or}\quad
R=24.1\,GeV^{-1}=4.75\,10^{-13}\mbox{cm}.\label{scale}
\ee
It turned out the fundamental scale has a clear physical meaning:
This scale corresponds to the scale of distances where the strong
Yukawa forces in strength come down to the electromagnetic forces
\cite{1}. After that we have built the Kaluza-Klein tower of KK
excitations by the formula 
\be
M_n=2\sqrt{m_p^2+\frac{n^2}{R^2}},\quad (n=1,2,3,\ldots)\label{KK}
\ee
and compared it with the observed irregularities in the mass spectrum
of diproton system. The result of the comparison has been
presented in extended Table 1 of ref. \cite{2}. It was established
that the Kaluza-Klein tower of KK excitations built by the calculated
fundamental scale was in a good correspondence with the
experimentally observed picture of irregularities in the mass
spectrum of two-nucleon system.

Now, let us build the Kaluza-Klein tower of KK excitations for
two-pion system by the formula 
\be
M_n^{\pi^1\pi^2} = \sqrt{m_{\pi^1}^2+\frac{n^2}{R^2}} +
\sqrt{m_{\pi^2}^2+\frac{n^2}{R^2}},\quad
(n=1,2,3,\ldots),\label{KKpi}
\ee
where $\pi^i(i=0,+,-)=\pi^0,\pi^+,\pi^-$ and $R$ is the same
fundamental scale (Eq. \ref{scale}) calculated early from the
analysis of nucleon-nucleon dynamics at low energies. Kaluza-Klein
tower such built is shown in Table 1 where the comparison with
experimentally observed mass spectrum of two-pion system is also
presented.

We have used Review of Particle Physics \cite{3} and recent review
article of Crystal Barrel Collaboration \cite{4} (see also many
references therein) where the experimental data on mass spectrum of
the resonance states of two-pion system have been extracted from. As
it is seen from Table 1, there is a quite remarkable correspondence
of the calculated KK excitations for two-pion system with the
experimentally observed mass spectrum of the resonance states of
two-pion system, which we consider as an additional strong evidence
of Kaluza-Klein picture of the world. 

Actually, we also see that there are three empty cells in the Table:
$M_1(282-291)$, $M_2(317-325)$ and $M_{13}(1112-1114)$. We did not
find an experimental confirmation of these states. Probably such
two-pion resonance states exist but we don't know these data. Tnat's
why any experimental information in this respect would gratefully be 
acknowledged.\footnote{Recently we found the references
\cite{9,10,11,12} (some of them is an old enough) which allowed us to
fill the cells 1 and 2.}

\begin{table}[tbp]
\begin{center}
\caption{Kaluza-Klein tower of KK excitations of two-pion system
and experimental data.}
\vspace{5mm}
{\large
\begin{tabular}{|c|c|c|c|c|}\hline   
 n & $ M_n^{\pi^0\pi^0}MeV $ & $ M_n^{\pi^0\pi^\pm}MeV $ & $
 M_n^{\pi^\pm\pi^\pm}MeV $ & $ M_{exp}^{\pi\pi}\,MeV $  \\
 \hline   
1  & 282.41  & 286.80  & 291.21   & $\sim$ 300  \\
2  & 316.87  & 320.80  & 324.73   & 322 $\pm$ 8  \\
3  & 367.18  & 370.58  & 373.98   & 370 -- i356   \\
4  & 427.78  & 430.71  & 433.64   & 430 -- i325   \\
5  & 494.92  & 497.45  & 499.99   & 506 $\pm$ 10  \\
6  & 566.26  & 568.48  & 570.70   & 585 $\pm$ 20  \\
7  & 640.41  & 642.38  & 644.34   & 650 -- i370   \\
8  & 716.50  & 718.26  & 720.01   & 732 -- i123   \\
9  & 793.96  & 795.55  & 797.13   & 780 $\pm$ 30  \\
10 & 872.44  & 873.88  & 875.33   & 870 -- i370   \\
11 & 951.68  & 953.00  & 954.32   & 955 $\pm$ 10  \\
12 & 1031.50 & 1032.72 & 1033.94  & 1015 $\pm$ 15 \\
13 & 1111.78 & 1112.92 & 1114.05  &   \\
14 & 1192.43 & 1193.49 & 1194.55  & 1165 $\pm$ 50   \\
15 & 1273.38 & 1274.37 & 1275.36  & 1275.4 $\pm$ 1.2  \\
16 & 1354.57 & 1355.50 & 1356.43  & 1359 $\pm$ 40  \\
17 & 1435.96 & 1436.84 & 1437.72  & 1434 $\pm$ 18  \\
18 & 1517.53 & 1518.36 & 1519.19  & 1522 $\pm$ 25  \\
19 & 1599.24 & 1600.02 & 1600.81  & 1593 $\pm$ $8\,^{+\,29}_{-\,47}$
\\
20 & 1681.07 & 1681.82 & 1682.57  & 1678 $\pm$ 12  \\
21 & 1763.00 & 1763.72 & 1764.43  & 1768 $\pm$ 21  \\
22 & 1845.03 & 1845.71 & 1846.40  & 1854 $\pm$ 20  \\
23 & 1927.14 & 1927.79 & 1928.45  & 1921 $\pm$ 8  \\
24 & 2009.32 & 2009.94 & 2010.57  & 2010 $\pm$ 60  \\
25 & 2091.56 & 2092.16 & 2092.76  & 2086 $\pm$ 15  \\
26 & 2173.85 & 2174.43 & 2175.01  & 2175 $\pm$ 20  \\
27 & 2256.19 & 2256.75 & 2257.31  & $\sim$ 2250  \\
28 & 2338.58 & 2339.12 & 2339.66  & $\sim$ 2330  \\
29 & 2421.01 & 2421.53 & 2422.05  & 2420 $\pm$ 30  \\
30 & 2503.47 & 2503.97 & 2504.48  & 2510 $\pm$ 30  \\ \hline
\end{tabular}}
\end{center}
\end{table}

Some known experimental information concerning resonance states in
two-pion system is collected in separate tables: Table 2 -- Table 22. 

We can learn from these tables that many different two-pion
resonances with the different quantum numbers may occupy one and the
same storey in KK tower. This is a peculiarity of the systematics
provided by Kaluza-Klein picture. Kaluza-Klein scenario in the
considered simplest case  predict only masses of KK excitations and
do not give any information on quantum numbers of corresponding
resonance states. The later information on quantum numbers
exceptionally depends on the physical process and the details of a
dynamics of the given physical process where KK excitations have been
observed. In particular, this also concerns the details (geometrical
structures and shapes) of generic compact internal extra
space\footnote{In the case of generic compact internal extra space KK
excitations are generally degenerated.}.

Of course, we would like to especially emphasize with a pleasure that
$\sigma$-meson ($M_\sigma \simeq 430\,MeV$, see discussion in
\cite{4}),
$f_2(0^+2^{++})$-mesons ($M_{f_2} = 1272\pm 8 \,MeV$ \cite{7} and
$M_{f_2} = 2175\pm 20 \,MeV$ \cite{8}) investigated by IHEP group
under direction of Yu.D.~Prokoshkin accurately agree with the
calculated values and excellently incorporated in the scheme of
systematics provided by Kaluza-Klein picture. 

\section{Conclusion}

The central point of  Kaluza-Klein approach is related to the
existence of a new fundamental scale characterizing a size of compact
internal extra space. In previous article \cite {1} we have
calculated this fundamental scale and shown that geniusly
simple formula (\ref{KK}) provided by Kaluza-Klein approach gives an
excellent description for the mass spectrum of two-nucleon system
\cite{2}. It has also been established that the experimental data
obtained at low energies where the nucleon-nucleon dynamics has been
studied reveal a special sort of (super)symmetry between
fermionic(dibaryonic) and bosonic states predicted by Kaluza-Klein
scenario.

In this note we have presented additional arguments in favour of
Kaluza and Klein picture of the world. In fact, we have shown that
formula (\ref{KKpi}) provided by Kaluza-Klein approach with the
fundamental scale early calculated (Eq. \ref{scale}) gives an
excellent description for the mass spectrum of two-pion system.

Of course, it would be very desirable to state new experiments to
search a further justification of the systematics provided by Kaluza
and Klein picture of the world. We believe that  this is a quite
promising subject of the investigations in particle and
nuclear physics.

\begin{table}[hbt]
\begin{center}
\caption{$M_{1}(282-291)$--Storey.}
\vspace{5mm}
\begin{tabular}{|c|c|c|c|c|}\hline   
$R(I^GJ^{PC})$ & $ M_R \, MeV $ & $ \Gamma_R \, MeV $ & Reaction &
Collab. \\ \hline   
 & $\sim$ 300 & $\sim$ 300 & $ \Upsilon(3S)
\rightarrow \pi^+\pi^-\Upsilon(2(1)S)$ & CLEO 91 \\ \hline
\end{tabular}
\end{center}
\end{table}

\begin{table}[hbt]
\begin{center}
\caption{$M_{2}(317-325)$--Storey.}
\vspace{5mm}
\begin{tabular}{|c|c|c|c|c|}\hline   
$R(I^GJ^{PC})$ & $ M_R \, MeV $ & $ \Gamma_R \, MeV $ & Reaction &
Collab.  \\ \hline   
``ABC"$(0^+0^{++})$ & 310 $\pm$ 10 & $\sim$ 25 & $ pd
\rightarrow He^3(\pi\pi)^0 $ & \cite{10} 60 \\
 & 322 $\pm$ 8 & $\leq$ 20 & $ \gamma p
\rightarrow p``ABC"$ & \cite{11} 62 \\
 & 300--365 & 45--75 & $ dp \rightarrow He^3(\pi\pi)^0$ & \cite{12}
 73 \\ \hline
\end{tabular}
\end{center}
\end{table}

\begin{table}[htb]
\begin{center}
\caption{$M_{11}(952-954)$--Storey.}
\vspace{5mm}
\begin{tabular}{|c|c|c|c|c|}\hline   
$R(I^GJ^{PC})$ & $ M_R \, MeV $ & $ \Gamma_R \, MeV $ & Reaction &
Collab. \\ \hline   
$f_0(0^+0^{++})$ & 955 $\pm$ 10 & 240 $\pm$ 60 & $ pp
\rightarrow pp\pi^0\pi^0 $ & GAM2 97 \\ \hline
\end{tabular}
\end{center}
\end{table}

\vspace{-1.5cm}
\begin{table}[htb]
\begin{center}
\caption{$M_{12}(1032-1034)$--Storey.}
\vspace{5mm}
\begin{tabular}{|c|c|c|c|c|}\hline   
$R(I^GJ^{PC})$ & $ M_R \, MeV $ & $ \Gamma_R \, MeV $ & Reaction &
Collab. \\ \hline   
$f_0(0^+0^{++})$ & 1015 $\pm$ 15 & 86 $\pm$ 16 & COMPILATION & RVUE
98 \\
 & $\sim$ 1015 & $\sim$ 30 & $ \pi\pi \rightarrow \pi\pi, K\bar K$ &
 RVUE 95 \\\hline
\end{tabular}
\end{center}
\end{table}

\vspace{-1.5cm}
\begin{table}[htb]
\begin{center}
\caption{$M_{14}(1192-1195)$--Storey.}
\vspace{5mm}
\begin{tabular}{|c|c|c|c|c|}\hline   
$R(I^GJ^{PC})$ & $ M_R \, MeV $ & $ \Gamma_R \, MeV $ & Reaction &
Collab. \\ \hline   
$f_0(0^+0^{++})$ & 1165 $\pm$ 50 & 460 $\pm$ 40 & $ \pi^{-}p
\rightarrow \pi^0\pi^0 n $ & RVUE 95 \\ \hline
\end{tabular}
\end{center}
\end{table}

%\newpage
\vspace{-1.5cm}
\begin{table}[htb]
\begin{center}
\caption{$M_{15}(1273-1275)$--Storey.}
\vspace{5mm}
\begin{tabular}{|c|c|c|c|r|}\hline   
$R(I^GJ^{PC})$ & $ M_R \, MeV $ & $ \Gamma_R \, MeV $ & Reaction &
Collab.\ \ \\ \hline   
$f_2(0^+2^{++})$ & 1275 $\pm$ 13 & 173 $\pm$ 53 & $\pi^+n \rightarrow
p\pi^+\pi^-$ & HBC 70 \\
 & 1276 $\pm$ 7 &        & $e^+e^- \rightarrow e^+e^{-}\pi^{+}\pi^-$
 & DLCO 84 \\
 & 1274 $\pm$ 5 &        & $J/\psi \rightarrow \gamma\pi^{+}\pi^-$ &
 DM2 87 \\
 & 1272 $\pm$ 8 & 192 $\pm$ 5 & $\pi^-p \rightarrow n\pi^{0}\pi^0$ &
 GAM2 94 \\
 & 1275.4 $\pm$ 1.2 & $156.9\,^{+\,3.8}_{-\,1.3}$ & AVERAGE & PDG 00
 \\ \hline
\end{tabular}
\end{center}
\end{table}

%\vspace{-1.5cm}
\begin{table}[htb]
\begin{center}
\caption{$M_{16}(1355-1356)$--Storey.} 
\vspace{5mm}
\begin{tabular}{|c|c|c|c|c|}\hline   
$R(I^GJ^{PC})$ & $ M_R \, MeV $ & $ \Gamma_R \, MeV $ & Reaction &
Collab. \\ \hline   
$f_0(0^+0^{++})$ & 1315 $\pm$ 50 & 255 $\pm$ 60 & $ pp \rightarrow
pp\pi^0\pi^0$ & GAM4 99 \\
$\rho(1^+1^{--})$ & 1348 $\pm$ 33 & 275 $\pm$ 10 & $\bar np
\rightarrow \pi^{+}\pi^{+}\pi^-$ & OBLX 98 \\
$\rho(1^+1^{--})$ & 1359 $\pm$ 40 & 310 $\pm$ 40 & $\bar pp
\rightarrow \pi^{+}\pi^{-}\pi^0$ & OBLX 97 \\
$\rho(1^+1^{--})$ & $1370\,^{+\,90}_{-\,70}$ &  & $ e^{+}e^-
\rightarrow \pi^{+}\pi^-$ &  RVUE 97 \\ \hline
\end{tabular}
\end{center}
\end{table}

%\vspace{-1.5cm}
\begin{table}[htb]
\begin{center}
\caption{$M_{17}(1435-1438)$--Storey.} 
\vspace{5mm}
\begin{tabular}{|c|c|c|c|r|}\hline   
$R(I^GJ^{PC})$ & $ M_R \, MeV $ & $ \Gamma_R \, MeV $ & Reaction &
Collab.\ \ \\ \hline   
$f_0(0^+0^{++})$ & 1420 $\pm$ 20 & 460 $\pm$ 50 & $ pp \rightarrow
pp\pi^+\pi^-$ & SPEC 86 \\
 & 1434$\pm$18$\pm$9 & 173$\pm$32$\pm$6 & $ D_s^+ \rightarrow
\pi^-\pi^+\pi^+$ & E791 01 \\
$f_2(0^+2^{++})$ & 1421 $\pm$ 5 & 30 $\pm$ 9 & $J/\psi
\rightarrow \gamma\pi^{+}\pi^-$ & DM2 87 \\
 & 1480 $\pm$ 50 & 150 $\pm$ 50 & $ pp
\rightarrow pp\pi^{+}\pi^{-}$ & SPEC 86 \\
$\rho(1^+1^{--})$ & 1424 $\pm$ 25 & 269 $\pm$ 31 & $ e^{+}e^-
\rightarrow \pi^{+}\pi^-$ & DM2 89 \\ \hline
\end{tabular}
\end{center}
\end{table}

\begin{table}[htb]
\begin{center}
\caption{$M_{18}(1518-1520)$--Storey.} 
\vspace{5mm}
\begin{tabular}{|c|c|c|c|r|}\hline   
$R(I^GJ^{PC})$ & $ M_R \, MeV $ & $ \Gamma_R \, MeV $ & Reaction &
Collab.\ \ \\ \hline   
$f_0(0^+0^{++})$ & 1522 $\pm$ 25 & 108 $\pm$ 33 & $\bar np
\rightarrow \pi^{+}\pi^{+}\pi^-$ & OBLX 98 \\
 & 1497 $\pm$ 30 & 199 $\pm$ 30 & $ pp \rightarrow
pp\pi^{+}\pi^-$ & OMEG 95 \\
 & 1502 $\pm$ 10 & 131 $\pm$ 15 & $ pp
\rightarrow pp\pi^{+}\pi^{-}$ & OMEG 99 \\
 & 1530 $\pm$ 45 & 160 $\pm$ 50 & $ pp
\rightarrow pp\pi^{0}\pi^0$ & GAM4 99 \\
 & 1580 $\pm$ 80 & 280 $\pm$ 100  & $\pi^{-}p
\rightarrow \pi^{0}\pi^{0}n$  &  GAM4 98 \\
$f_2(0^+2^{++})$ & 1507 $\pm$ 15 & 130 $\pm$ 20  & $\bar pp
\rightarrow \pi^{+}\pi^{-}\pi^0$  &  OBLX 97 \\
 & 1540 $\pm$ 15 & 132 $\pm$ 37  & $\bar np
\rightarrow \pi^{+}\pi^{+}\pi^{-}$  &  OBLX 92 \\
$f_2'(0^+2^{++})$ & 1502 $\pm$ 25 & 165 $\pm$ 42 & $\pi^{-}p
\rightarrow \pi^{+}\pi^{-}n$ & OMEG 79 \\\hline
\end{tabular}
\end{center}
\end{table}

\begin{table}[htb]
\begin{center}
\caption{$M_{19}(1599-1601)$--Storey.} 
\vspace{5mm}
\begin{tabular}{|c|c|c|c|r|}\hline   
$R(I^GJ^{PC})$ & $ M_R \, MeV $ & $ \Gamma_R \, MeV $ & Reaction &
Collab.\ \ \\ \hline   
$\pi_1(1^-1^{-+})$ & 1593 $\pm$ $8\,^{+\,29}_{-\,47}$ & 168 $\pm
60\,^{+\,150}_{-\,12}$ & $ \pi^{-}p \rightarrow
\pi^+\pi^-\pi^{-}p$ & MPS 98 \\
$f_0(0^+0^{++})$ & 1580 $\pm$ 80 & 280 $\pm$ 100  & $\pi^{-}p
\rightarrow \pi^{0}\pi^{0}n$  &  GAM4 98 \\ \hline
\end{tabular}
\end{center}
\end{table}

\vspace{-1cm}
\begin{table}[htb]
\begin{center}
\caption{$M_{20}(1681-1683)$--Storey.} 
\vspace{5mm}
\begin{tabular}{|c|c|c|c|r|}\hline   
$R(I^GJ^{PC})$ & $ M_R \, MeV $ & $ \Gamma_R \, MeV $ & Reaction &
Collab.\ \ \\ \hline   
$\rho_3(1^+3^{--})$ & 1677 $\pm$ 14 & 246 $\pm$ 37 & $\pi^{-}p
\rightarrow 2\pi p$ & OMEG 81 \\
 & 1679 $\pm$ 11 & 116 $\pm$ 30 & $\pi^{+}p \rightarrow
\pi^{+}\pi^{-}n$ & HBC 78 \\
 & 1678 $\pm$ 12 & 162 $\pm$ 50 & $\pi^{-}p
\rightarrow p3\pi$ & CIBS 77 \\
 & 1690 $\pm$ 7 & 167 $\pm$ 40 & $\pi^{+}n
\rightarrow \pi^{+}\pi^{-}p$ & DBC 74 \\
 & 1693 $\pm$ 8 & 200 $\pm$ 18  & $\pi^{-}p
\rightarrow \pi^{+}\pi^{-}n$  &  ASPK 74 \\
 & 1678 $\pm$ 12 & 156 $\pm$ 36 & $ \pi^{+}N $ & DBC 71 \\
 & 1692 $\pm$ 12 & 240 $\pm$ 30  & $\pi^{-}p
\rightarrow \pi^{+}\pi^{-}n$  & RVUE 75 \\
 & 1650 $\pm$ 35 & 180 $\pm$ 30  & $\pi^{-}p
\rightarrow N2\pi $  &  HBC 70 \\ 
 & 1687 $\pm$ 21 & $267\,^{+\,72}_{-\,46}$   & $\pi^{-}p, \pi^{+}d$
 &
 HDBC 70 \\ 
 & 1683 $\pm$ 13 & 188 $\pm$ 49 & $\pi^{+}d$  & DBC 68 \\ 
 & 1670 $\pm$ 30 & 180 $\pm$ 40 & $\pi^{+}d, \pi^{-}p$ & HBC 65 \\
 & 1688.8 $\pm$ 2.1 & 186 $\pm$ 14 & AVERAGE & PDG 00 \\ \hline
\end{tabular}
\end{center}
\end{table}

\vspace{-1cm}
\begin{table}[htb]
\begin{center}
\caption{$M_{21}(1763-1764)$--Storey.} 
\vspace{5mm}
\begin{tabular}{|c|c|c|c|r|}\hline   
$R(I^GJ^{PC})$ & $ M_R \, MeV $ & $ \Gamma_R \, MeV $ & Reaction &
Collab. \ \ \\ \hline   
$\rho(1^+1^{--})$ & $1780\,^{+\,37}_{-\,29}$ & 275 $\pm$ 45 & $\bar
pn
\rightarrow \pi^{-}\pi^0\pi^0$ & CBAR 97 \\
 & 1730 $\pm$ 30 & 400 $\pm$ 100 & $e^{+}e^- \rightarrow
\pi^{+}\pi^{-}$ & RVUE 94 \\
 & 1768 $\pm$ 21 & 224 $\pm$ 22 & $e^{+}e^- \rightarrow
\pi^{+}\pi^{-}$ & DM2 89 \\
 & 1745.7 $\pm$ 91.9 & 242.5 $\pm$ 163 & $e^{+}e^- \rightarrow
\pi^{+}\pi^{-}$ & RVUE 89 \\
$f_0(0^+0^{++})$ & $1740\,^{+\,30}_{-\,25}$ & $120\,^{+\,50}_{-\,40}$
&
$J/\psi \rightarrow \gamma(\pi^{+}\pi^{-}\pi^{+}\pi^{-} )$ & BES 00
\\
 & 1750 $\pm$ 20 & 160 $\pm$ 30 & $ pp \rightarrow pp\pi^{+}\pi^{-} $
 & OMEG 99 \\
 & 1750 $\pm$ 15 & 160 $\pm$ 40  & $J/\psi \rightarrow
 \gamma(\pi^{+}\pi^{-}\pi^{+}\pi^{-} )$ & MRK3 95 \\
 & 1750 $\pm$ 30 & 250 $\pm$ 140  & COMPILATION & RVUE 98 \\ \hline
\end{tabular}
\end{center}
\end{table}

\begin{table}[htb]
\begin{center}
\caption{$M_{22}(1845-1846)$--Storey.} 
\vspace{5mm}
\begin{tabular}{|c|c|c|c|r|}\hline   
$R(I^GJ^{PC})$ & $ M_R \, MeV $ & $ \Gamma_R \, MeV $ & Reaction &
Collab. \ \ \\ \hline   
$\eta_2(0^+2^{-+})$ & 1840 $\pm$ 25 & 200 $\pm$ 40 & $ pp \rightarrow
pp2(\pi^+\pi^-)$ & OMEG 97 \\  
 & 1881 $\pm$ 32 $\pm$ 40 & 221 $\pm$ 92 $\pm$ 44 & $ e^{+}e^-
 \rightarrow e^{+}e^-\eta\pi^0\pi^0 $ & CBAL 92 \\ 
  & 1840 $\pm$ 15 & 170 $\pm$ 40 & $ J/\psi \rightarrow
\gamma\eta\pi^+\pi^-$ & BES 99 \\ 
 & 1854 $\pm$ 20 & 202 $\pm$ 30 & AVERAGE & PDG 00 \\ \hline
\end{tabular}
\end{center}
\end{table}

\vspace{-1.5cm}
\begin{table}[htb]
\begin{center}
\caption{$M_{23}(1927-1928)$--Storey.} 
\vspace{5mm}
\begin{tabular}{|c|c|c|c|r|}\hline   
$R(I^GJ^{PC})$ & $ M_R \, MeV $ & $ \Gamma_R \, MeV $ & Reaction &
Collab. \ \ \\ \hline 
$X(0^+2^{++})$ & 1920 $\pm$ 10 & 90 $\pm$ 20 & $\pi^{-}p \rightarrow
\omega\omega n$ & VES 92 \\
 & 1924 $\pm$ 14 & 91 $\pm$ 50 & $\pi^{-}p \rightarrow
\omega\omega n$ & GAM2 90 \\ 
 & 1921 $\pm$ 8 & 90 $\pm$ 19 & AVERAGE & PDG 00 \\ 
$f_2(0^+2^{++})$ & 1918 $\pm$ 12 & 390 $\pm$ 60 & $ pp \rightarrow
pp2(\pi^+\pi^-)$ & OMEG 95 \\
 & 1940 $\pm$ 50 & $380\,^{+\,120}_{-\,90}$ & $J/\psi \rightarrow
 \gamma(\pi^{+}\pi^{-}\pi^{+}\pi^{-} )$ & BES 00 \\
 & 1960 $\pm$ 30 & 460 $\pm$ 40 & $ pp \rightarrow
pp2(\pi^+\pi^-)$ & OMEG 97 \\ 
$f_4(0^+4^{++})$ & 1935 $\pm$ 13 & 263 $\pm$ 57 & $\pi^{-}p
\rightarrow n2\pi$ & OMEG 79 \\ \hline
\end{tabular}
\end{center}
\end{table}

\vspace{-1.5cm}
\begin{table}[htb]
\begin{center}
\caption{$M_{24}(2009-2011)$--Storey.} 
\vspace{5mm}
\begin{tabular}{|c|c|c|c|r|}\hline   
$R(I^GJ^{PC})$ & $ M_R \, MeV $ & $ \Gamma_R \, MeV $ & Reaction &
Collab. \ \ \\ \hline   
$f_2(0^+2^{++})$ & $\sim$1996 & $\sim$134 & $\bar pp
\rightarrow \pi\pi$ & RVUE 94 \\
 & $\sim$1990 & $\sim$100 & $\bar pp \rightarrow
\pi\pi$ & RVUE 94 \\
 & 2010 $\pm$ 60 & 240 $\pm$ 100 & $\pi^{-}p
\rightarrow \pi^0\pi^0 n$ & GAM4 98 \\ 
$f_0(0^+0^{++})$ & 2010 $\pm$ 60 & 240 $\pm$ 100  & $\pi^{-}p
\rightarrow \pi^{0}\pi^{0}n$  &  GAM4 98 \\
 & 2020 $\pm$ 35 & 410 $\pm$ 50 & $pp \rightarrow pp2(\pi^+\pi^-)$ &
 OMEG 97 \\
$f_4(0^+4^{++})$ & 1998 $\pm$ 15 & 395 $\pm$ 40 & $\pi^{-}p
\rightarrow \pi^{0}\pi^{0}n$  & GAM4 98 \\
 & 2032 $\pm$ 30 & 304 $\pm$ 60  & $J/\psi \rightarrow
 \gamma\pi^+\pi^-$  & DM2 87 \\
 & 2020 $\pm$ 20 & 240 $\pm$ 40  & $\pi^{-}p
\rightarrow n2\pi $  &  GAM2 84 \\ 
 & 2015 $\pm$ 28 & $186\,^{+\,103}_{-\,58}$ & $\pi^{+}p \rightarrow
 \Delta^{++}\pi^0\pi^0$  & STRC 82 \\ 
 & 2020 $\pm$ 30 & 180 $\pm$ 60 & $\pi^{-}p
\rightarrow n2\pi^0 $  & NICE 75 \\ 
 & $\sim$2000 & $\sim$170 & $N\bar N \rightarrow \pi\pi$ & RVUE 98 \\
 & $\sim$2010 & $\sim$200 & $N\bar N \rightarrow \pi\pi$ & RVUE 97 \\ 
 & 2034 $\pm$ 11 & 222 $\pm$ 19 & AVERAGE & PDG 00 \\ 
$\rho_3(1^+3^{--})$ & $\sim$2007 & $\sim$287 & $\bar pp \rightarrow
\pi\pi$ & RVUE 94 \\ \hline
\end{tabular}
\end{center}
\end{table}

\begin{table}[htb]
\begin{center}
\caption{$M_{25}(2091-2093)$--Storey.}
\vspace{5mm}
\begin{tabular}{|c|c|c|c|c|}\hline   
$R(I^GJ^{PC})$ & $ M_R \, MeV $ & $ \Gamma_R \, MeV $ & Reaction &
Collab. \\ \hline   
$f_4(0^+4^{++})$ & 2086 $\pm$ 15 & 210 $\pm$ 63 & $ J/\psi
\rightarrow \gamma\pi^+\pi^-$ & MRK3 87 \\
$\rho_3(1^+3^{--})$ & $\sim$2090 & $\sim$ 60 & $ \bar pp
\rightarrow \pi\pi$ & RVUE 94 \\\hline
\end{tabular}
\end{center}
\end{table}

%\vspace{-1.5cm}
\begin{table}[htb]
\begin{center}
\caption{$M_{26}(2174-2175)$--Storey.} 
\vspace{5mm}
\begin{tabular}{|c|c|c|c|c|}\hline   
$R(I^GJ^{PC})$ & $ M_R \, MeV $ & $ \Gamma_R \, MeV $ & Reaction &
Collab. \\ \hline 
$f_2(0^+2^{++})$ & 2175 $\pm$ 20 & 150 $\pm$ 35 & $pp \rightarrow
pp2\eta $ & GAM4 95 \\
 & $\sim$ 2170 & $\sim$250 & $\bar pp \rightarrow
\pi\pi$ & RVUE 80 \\ 
$\rho(1^+1^{--})$ & $\sim$ 2170 & $\sim$ 250  & $\bar pp \rightarrow
\pi\pi$ & RVUE 80 \\ 
 & $\sim$ 2191 & $\sim$ 296 & $\bar pp \rightarrow \pi\pi$ & RVUE 94
 \\
 & 2153 $\pm$ 37 & 389 $\pm$ 79 & $e^{+}e^- \rightarrow \pi^+\pi^-$ &
 RVUE 91 \\
$\rho_3(1^+3^{--})$ & $\sim$ 2150 & $\sim$ 200 & $\bar pp \rightarrow
\pi\pi$ & CNTR 77 \\ \hline
\end{tabular}
\end{center}
\end{table}

%\vspace{-1.5cm}
\begin{table}[htb]
\begin{center}
\caption{$M_{27}(2256-2257)$--Storey.} 
\vspace{5mm}
\begin{tabular}{|c|c|c|c|r|}\hline   
$R(I^GJ^{PC})$ & $ M_R \, MeV $ & $ \Gamma_R \, MeV $ & Reaction &
Collab. \ \ \\ \hline 
$f_2(0^+2(4)^{++})$ & 2246 $\pm$ 36 &  & $J/\psi \rightarrow
\gamma\pi^0\pi^0 $ & BES 98 \\
$\rho_3(1^+3^{--})$ & $\sim$ 2250 & $\sim$ 250 & $\bar pp \rightarrow
\pi\pi$ & RVUE 80 \\ 
 & $\sim$ 2232 & $\sim$ 220  & $\bar pp \rightarrow
\pi\pi$ & RVUE 94 \\ 
$\rho_5(1^+5^{--})$ & $\sim$ 2250 & $\sim$ 300 & $\bar pp \rightarrow
\pi\pi$ & RVUE 80 \\ \hline
\end{tabular}
\end{center}
\end{table}

\newpage
\begin{table}[htb]
\begin{center}
\caption{$M_{28}(2339-2340)$--Storey.} 
\vspace{5mm}
\begin{tabular}{|c|c|c|c|c|}\hline   
$R(I^GJ^{PC})$ & $ M_R \, MeV $ & $ \Gamma_R \, MeV $ & Reaction &
Collab. \\ \hline 
$f_0(0^+0^{++})$ & $\sim$ 2321 & $\sim$ 223 & $\bar pp \rightarrow
\pi\pi $ & RVUE 94 \\
$f_4(0^+4^{++})$ & $\sim$ 2314 & $\sim$ 278 & $\bar pp \rightarrow
\pi\pi$ & RVUE 94 \\ 
 & $\sim$ 2300 & $\sim$ 200  & $\bar pp \rightarrow
\pi\pi$ & RVUE 80 \\ 
 & $\sim$ 2330 & $\sim$ 300 & $\bar pp \rightarrow \pi^0\pi^0$ & OSPK
 78
 \\
 & $\sim$ 2310 & $\sim$ 200 & $ \bar pp \rightarrow \pi\pi$ &
 CNTR 77 \\
$\rho_5(1^+5^{--})$ & $\sim$ 2303 & $\sim$ 169 & $\bar pp \rightarrow
\pi\pi$ & RVUE 94 \\
 & $\sim$ 2300 & $\sim$ 250 & $\bar pp \rightarrow
\pi\pi$ & RVUE 80 \\\hline
\end{tabular}
\end{center}
\end{table}

\vspace{-1.5cm}
\begin{table}[htb]
\begin{center}
\caption{$M_{29}(2421-2422)$--Storey.}
\vspace{5mm}
\begin{tabular}{|c|c|c|c|c|}\hline   
$R(I^GJ^{PC})$ & $ M_R \, MeV $ & $ \Gamma_R \, MeV $ & Reaction &
Collab.  \\ \hline   
$f_6(0^+6^{++})$ & 2420 $\pm$ 30 & 270 $\pm$ 60 & $ \pi^{-}p
\rightarrow \pi^0\pi^0 n$ & GAM4 98 \\ \hline
\end{tabular}
\end{center}
\end{table}

\vspace{-1.5cm}
\begin{table}[htb]
\begin{center}
\caption{$M_{30}(2503-2504)$--Storey.}
\vspace{5mm}
\begin{tabular}{|c|c|c|c|c|}\hline   
$R(I^GJ^{PC})$ & $ M_R \, MeV $ & $ \Gamma_R \, MeV $ & Reaction &
Collab. \\ \hline   
$f_6(0^+6^{++})$ & 2510 $\pm$ 30 & 240 $\pm$ 60 & $ \pi^{-}p
\rightarrow \pi^0\pi^0 n$ & GAM2 84 \\
$\rho_5(1^+5^{--})$ & $\sim$2480 & $\sim$ 210 & $ \bar pp
\rightarrow \pi\pi$ & CNTR 77 \\\hline
\end{tabular}
\end{center}
\end{table}


\begin{thebibliography}{**}
\bibitem{1}
A.A.~Arkhipov, hep-ph/0208215 (2002); preprint IHEP 2002-43,
Protvino, 2002,
available at http://dbserv.ihep.su/\~{}pubs/prep2002/ps/2002-43.pdf
\bibitem{2}
A.A.~Arkhipov, hep-ph/0302164 (2003).
\bibitem{3}
D.E.~Groom et al., {\it Review of Particle Physics}, Eur. Phys. J.
C{\bf 15} 405, 417-487 (2000).
\bibitem{4}
V.V.~Anisovich, hep-ph/0208123 (2002).
\bibitem{5}
V.A.~Rubakov, {\it Large and infinite extra dimensions}, Sov. J.
Uspekhi {\bf 171}, 913 (2001); e-print hep-ph/0104152.
\bibitem{6}
Yu.A.~Kubyshin, {\it Models with Extra Dimensions and Their
Phenomenology}, e-print hep-ph/0111027.
\bibitem{7}
A.A.~Kondashov, Yu.D.~Prokoshkin, Physics--Doklady, {\bf 336}, 613
(1994).
\bibitem{8}
Yu.D.~Prokoshkin, Physics--Doklady, {\bf 344}, 469 (1995).
\bibitem{9}
C.~Bebek et al. (CLEO Collab.), Phys. Rev. D{\bf 43}, 1448 (1991).
\bibitem{10}
A.~Abashian, N.E.~Booth, K.M.~Crowe, Phys. Rev. Lett. {\bf 5}, 258
(1960), {\it ibid.} {\bf 7}, 35 (1961).
\bibitem{11}
B.~Richter, Phys. Rev. Lett. {\bf 9}, 217 (1962).
\bibitem{12}
J.~Banaigs et al., Nucl. Phys. B{\bf 67}, 1 (1973).
\end{thebibliography}
\end{document}